\documentclass{sf2a-conf2016}
\usepackage{graphicx}
\usepackage{hyperref}
\usepackage[]{natbib}  
\usepackage{epstopdf}
\usepackage{todonotes}

\def\BibTeX{{\rm B\kern-.05em{\sc i\kern-.025em b}\kern-.08em
    T\kern-.1667em\lower.7ex\hbox{E}\kern-.125emX}}
\bibpunct{(}{)}{;}{a}{}{,}  
\newcommand{\kms}{{\mathrm{km~s^{-1}}}}

\newcommand{\teff}{{\mathrm{Teff}}}
\newcommand{\logg}{{\mathrm{\log g}}}

\begin{document}

\TitreGlobal{SF2A 2016}


\title{HD 30963: a new HgMn star}

\runningtitle{HD30963}

\author{R.Monier$^{1,}$}\address{LESIA, UMR 8109, Observatoire de Paris Meudon, Place J.Janssen, Meudon, France}\address{Lagrange, UMR 7293, Universite de Nice Sophia, Nice, France}
\author{M.Gebran}\address{Department of Physics and Astronomy, Notre Dame University - Louaize, PO Box 72, Zouk Mikael, Lebanon}
\author{F.Royer}\address{GEPI, UMR 8111, Observatoire de Paris Meudon, Place J.Janssen, Meudon, France}


\setcounter{page}{237}


\maketitle


\begin{abstract}
Using high dispersion high quality spectra of HD 30963 obtained with the echelle spectrograph SOPHIE at Observatoire de Haute Provence in November 2015, we show that this star, hitherto classified as a B9 III superficially normal star, is actually  a new Chemically Peculiar star of the HgMn type. Spectrum synthesis reveals large overabundances of Mn, Sr, Y, Zr , Pt and Hg and pronounced underabundances of He and Ni which are characteristic of HgMn stars. We therefore propose that this interesting object be reclassified as a B9 HgMn star.
\end{abstract}



\section{Introduction}
  HD 30963 (V =7.23 mag) ,currently assigned an B9 III spectral type, is one of the slowly rotating late B we are currently observing. This star has been little studied: one only finds 10 references in SIMBAD for HD 30963. The incentive of this project is to reclassify the late B stars in the northern hemisphere brighter than V=7.5 mag with low apparent projected velocities (less than 60 $\kms$). We have previously undertaken a similar survey of the slowly rotating early A stars,.whose results are published in \cite{Royer14}.
   An abundance analysis of high resolution well exposed spectra of these objects has sorted out the sample into 17 chemically normal stars (ie. whose abundances do not depart more than $\pm$ 0.20 dex from solar values), 12 spectroscopic binaries and 13 Chemically Peculiar stars (CP).
   Among the new CP stars, \cite{Monier15} and  \cite{Monier2016} have identified 5 new HgMn stars. 
    We present here new abundance determinations for HD 30963, a slowly rotating late B star and show  that this star is another new HgMn late B star.

\section{Observations and reduction}

HD 30963 has been observed on five consecutive days at Observatoire de Haute Provence (OHP) using SOPHIE in its high resolution mode (R=75000) in November and December 2015. A log of the observations appears in Tab.~\ref{tab1}.
Four 1800 s and one 2400 seconds exposures were obtained  with a signal-to-noise ratio varying from 78 up to 167.  
As we do not see conspicuous radial velocity variations on these spectra taken over five days, we have 
shifted all spectra onto a common wavelength scale (that of the first spectrum) and we have coadded the individual spectra into a mean spectrum..
Assuming a radius for HD 30963 of about 3 to 4 $R_{\odot}$ and using the $v_{e} \sin i$ of \cite{royerzorec}, we derive an upper limit of 4.12 days for the rotational period of HD 30963.
The observations we secured over 5 days separated by about 24 hours each therefore cover the rotational period of HD 30963. 

\begin{table}
\centering
\begin{tabular}{|l|l|l|p{3cm}|}
 \hline
  \multicolumn{3}{|c|}{\textbf{Observations log for HD 30963}}   \\ \hline
  \multicolumn{1}{|c|}{Date} &
  \multicolumn{1}{|c|}{UT start exposure} &
  \multicolumn{1}{|c|}{$\frac{S}{N}$}          \\ \hline
  2015-11-28 & 00:00:35.307 &    118    \\ \hline
 2015-11-28  & 23:44:20.941  &    78     \\ \hline
 2015-11-30  &00:25:54.328  &  140     \\ \hline
 2015-11-30  &23:24:27.855 &   101    \\ \hline
 2015:12:01  & 22:29:32.033 &    167    \\ \hline
\end{tabular}
\caption{Log of SOPHIE observations of HD 30963}\label{tab1}  
\end{table} 

\section{Reassigning a proper spectral type to HD 30963}

\begin{figure}[ht!]
 \centering
 \includegraphics[width=0.8\textwidth]{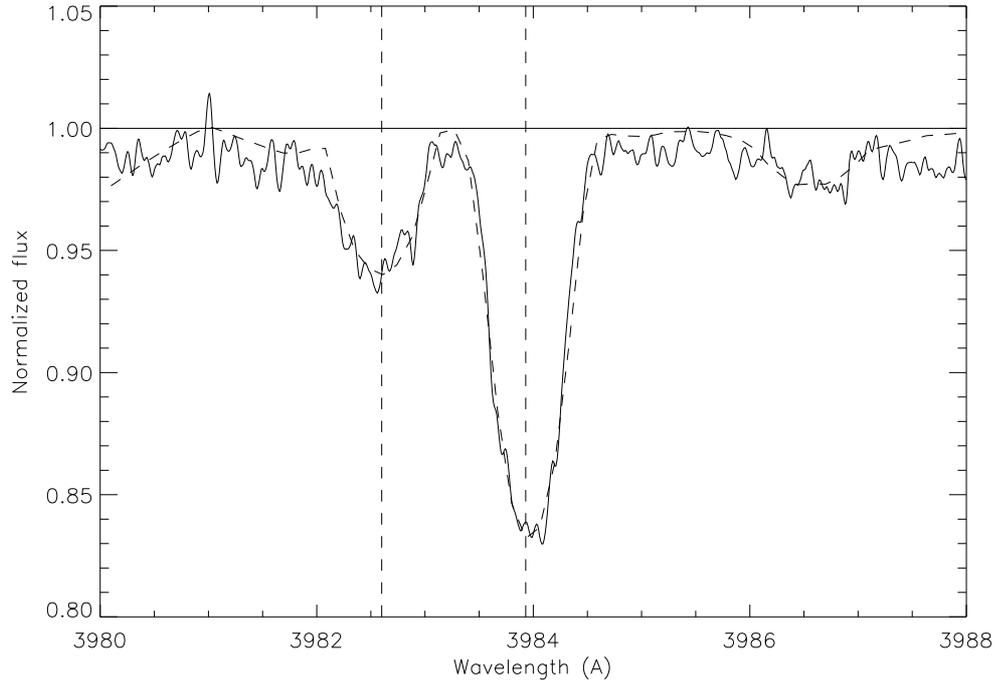}     
  \caption{Synthesis of the Hg II 3983.93 \AA\ line (observed: thick line, models: dashed lines)}
  \label{fig1}
\end{figure}

The following spectral regions have been used to readdress the spectral type of HD 30963. First, the red wing of $H_{\epsilon}$ from 3980 \AA\ up to 4000 \AA\ harbours the Hg II line at 3984 \AA\ and two Zr II and one Y II lines likely to be strengthened in a late B-type star of the HgMn type (Fig.~\ref{fig1}). Second, the region from 4470 \AA\ to 4490 \AA\ contains the Mg II triplet near 4481.15 \AA, the He I line at 4471.48 \AA\ and the Mn II line at 4478.64 \AA (Fig.~\ref{fig2}). We have also looked for peculiarities
in the Ni II line at 4067.04 \AA\ and the Sr II resonance line at 4077.70 \AA\ and in the Pt II line at 4514.17 \AA\ and the Ba II resonance line at 4554.01 \AA.
The eleven lines used are collected in Tab.~\ref{tab2} with their identifications and abundances (see Sec.~\ref{abun}) expressed as multiples of the corresponding solar abundances (labeled 
with the symbol $\odot$).
Their positions are marked by vertical lines in both figures (after a correction for a radial velocity).

\begin{table}
\centering
\begin{tabular}{|l|l|l|p{3cm}|}
 \hline
  \multicolumn{3}{|c|}{\textbf{Derived abundances for HD 30963}}   \\ \hline
  \multicolumn{1}{|c|}{Wavelengths (\AA)} &
  \multicolumn{1}{|c|}{Identification} &
  \multicolumn{1}{|c|}{Abundance}          \\ \hline
  3982.60    &   Y II    &      1000.0  $\odot$  \\ \hline
  3983.93    &  Hg II   &      150000.0 $\odot$  \\ \hline
  3990.96    &  Zr II    &     150.0 $\odot$   \\ \hline
  3998.82    &  Zr II    &    150.0 $\odot$    \\ \hline
  4067.04    & Ni II     &   0.10 $\odot$    \\ \hline
  4077.70    & Sr II     &   35.0 $\odot$       \\ \hline
  4471.48    & He I     &   0.20 $\odot$     \\ \hline
  4478.64    & Mn II   &   50.0 $\odot$      \\ \hline
  4481.15    & Mg II   &   2.50 $\odot$    \\ \hline
   4500.00   & Fe II   &    2.00 $\odot$     \\ \hline
   4514.17   & Pt II    &  2500.0 $\odot$   \\ \hline
   4554.01   & Ba II   & 10.00  $\odot$     \\ \hline
\end{tabular}
\caption{Abundances derived for the 11 lines used in this work.}  \label{tab2}
\end{table}  


\section{Abundance determinations}

\subsection{Model atmospheres and spectrum synthesis }

The effective temperature $\teff$ and surface gravity $\logg$ of HD 30963 were taken from \cite{huang} survey of field B stars. These authors have derived 
$\teff = 11476$ K and $\logg = 3.66 $ by adjusting the observed $H_{\gamma}$ profile of HD 30963 to grids of synthetic spectra.
A 72 layers plane parallel model atmosphere assuming radiative equilibrium and hydrostatic equilibrium has been first computed for these values of $\teff$ and $\logg$ 
 using the ATLAS9 code \citep{Kurucz92}, specifically the linux version using the new ODFs maintained by F. Castelli on her website. The linelist was built starting from Kurucz's (1992) gfhyperall.dat\footnote{http://kurucz.harvard.edu/linelists/} file which includes hyperfine splitting levels.
This first linelist was then upgraded using the NIST Atomic Spectra Database 
\footnote{http://physics.nist.gov/cgi-bin/AtData/linesform} and the VALD\footnote{http://vald.astro.uu.se/~vald/php/vald.php} database operated at Uppsala University \citep{kupka2000}.
A grid of synthetic spectra was then computed with SYNSPEC48 \citep{Hubeny92} to model the eleven lines listed in Table 1 adopting a null microturbulent velocity. 
The synthetic spectra were further convolved using the routine ROTIN3 provided along with SYNSPEC48 using $v_{e} \sin i = 37 $ $\kms$  \citep{royerzorec} and he appropriate FWHM for SOPHIE.
The unknown abundance $[\frac{X}{H}]$ was varied until minimisation of the chi-square between the observed and synthetic spectrum. 

\subsection{Evidence for strong excesses in Mn, Y, Zr, Pt and Hg}
\label{abun}
\begin{figure}[ht!]
 \centering
 \includegraphics[width=0.8\textwidth]{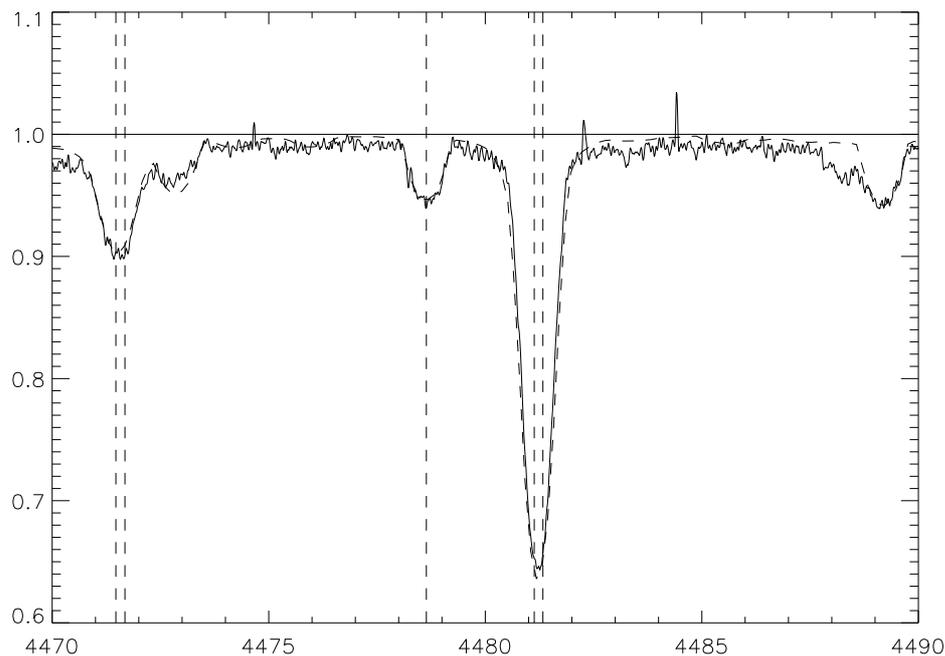}     
  \caption{Synthesis of the He I (4471.4 \AA), Mn II ( 4478.64 \AA) and Mg II (4481.15 \AA) lines}
  \label{fig2}
\end{figure}

We find that five elements, Mn, Y, Zr, Pt and Hg, have strong excesses, ie. larger than 50 times the solar values
The synthesis of the Mn II line at 4478.64 \AA\ in Fig.~\ref{fig1} yields an overabundance of 50 times the solar abundance. This is a substantial excess
and it may be revised to a lower value when we will model Mn II lines having atomic data for several isotopes and hyperfine structure.
The Hg overabundance is derived from the synthesis of 11 transitions next to 3983.84 \AA\ representing the hyperfine structure of various isotopes of Hg in a similar manner as \citep{Castelli04} 
 modeled HD 175640. The abundance that best fits the observed line profile is very large (150000 $\odot$) as can be seen in Fig.~\ref{fig2}.  This is one of the strongest Hg excess we found in the new HgMn stars \citep{Monier15,Monier2016}.

The abundance  of Zr derived from the 2 lines in Table 1 is about 150 $\odot$. The Yttrium abundance derived from Y II 3932.44 \AA\   is also very important (1000 $\odot$).
While modeling the Fe II lines in the 4500 \AA - 4600\AA\ region, we noticed a 1.5 \% feature at about 4514.15 \AA\ which we identify as one of the strongest Pt II line expected at 4514.17 \AA\ listed in VALD. No other "easier" identification (ie. in terms of a lighter element) can reproduce the observed line depth at this wavelength. We are currently looking for other lines of Pt II lines at shorter wavelengths to confirm the presence of platinum in HD 30963. 
 We also checked that a model computed for a solar platinum abundance  (and without the Pt II line) yields no absorption at all at 4514.17 \AA.  From the 4514.17 \AA\ line only, a provisional excess of platinum is derived to be about 2500 $\odot$.
 We also find considerable deficiencies in He (0.2 $\odot$) and Ni (0.1 $\odot$). The excesses in Mn, Y, Zr, Pt and Hg and the deficiencies in He and Ni are characteristic of an HgMn star.
\\

\section{Conclusions}

Whereas it was up to now classified as a normal late B9 III star, our analysis of HD 30963  shows that it has very peculiar over and underabundances. The overabundances in Mn, Sr, Y, Zr, Pt and Hg are characteristic of an HgMn star. It displays large overabundances of the Sr, Y and Zr triad which is however inverted compared to the solar system triad. The synthesis of the Hg II and Pt II lines reveals large overabundances of Pt and Hg. We are currently performing a detailed abundance analysis of HD 30963 to complement the first abundances presented here.

\begin{acknowledgements}
The authors acknowledge the efficient help of the night assistants at Observatoire de Haute Provence. They have used the NIST Atomic Spectra Database and the VALD database operated at Uppsala University (Kupka et al., 2000) to upgrade atomic data.
\end{acknowledgements}



%
\end{document}